\documentclass[preprint,review,3p, 12pt]{elsarticle}

\biboptions{numbers,sort&compress}

\usepackage[T1]{fontenc}

\usepackage[many]{tcolorbox}

\RequirePackage[colorlinks,citecolor=blue,urlcolor=blue,linkcolor=blue]{hyperref}
\RequirePackage{url}

\usepackage[colorinlistoftodos]{todonotes}

\journal{Current Opinion in Structural Biology}

\begin{document}
\begin{frontmatter}
	
\title{Minimal coarse-grained models for molecular self-organisation in biology}

\author[a1,a2]{Anne E. Hafner}
\author[a1,a2]{Johannes Krausser}
\author[a1]{An\dj ela \v{S}ari\'c \corref{cor1} }

\address[a1]{Department of Physics and Astronomy, Institute for the Physics of Living Systems, University College London, London WC1E 6BT, UK}
\address[a2]{These authors contributed equally.}

\cortext[cor1]{E-mail: a.saric@ucl.ac.uk}


\begin{abstract}
\noindent  

The molecular machinery of life is largely created via self-organisation of individual molecules into functional assemblies. Minimal coarse-grained models, where a whole macromolecule is represented by a small number of particles, can be of great value in identifying the main driving forces behind self-organisation in cell biology. Such models can incorporate data from both molecular and continuum scales, and their results can be directly compared to experiments. Here we review the state of the art of models for studying the formation and biological function of macromolecular assemblies in cells. We outline the key ingredients of each model and their main findings. We illustrate the contribution of this class of simulations to identifying the physical mechanisms behind life and diseases, and discuss their future developments.
\end{abstract}

\end{frontmatter}


\section*{Introduction}

Self-assembly of individual molecules into large-scale functional structures generates the molecular machinery of life~\cite{Robinson2007}. 
Such processes underlie the formation of cell membranes, protein filaments and networks, and drive the formation of three-dimensional genome structures. Many of these assemblies, such as protein filaments and lattices,  also function as efficient nanomachines that enable cells to sense, move, divide, and transport materials in and out of the cell~\cite{Bassereau2018}. To sustain life, the formation of macromolecular assemblies needs to be dynamic, reversible, and necessarily driven out of equilibrium. Hence, functional assemblies consume energy and continuously undergo turnover. Conversely, when the assembly in living organisms occurs irreversibly, without energy input, severe pathologies can occur. A prominent example of the latter is the assembly of amyloid fibrils and plaques involved in over 50 human diseases~\cite{Dobson2019}. 

\begin{figure*}
	\centering
	\includegraphics[width = \linewidth]{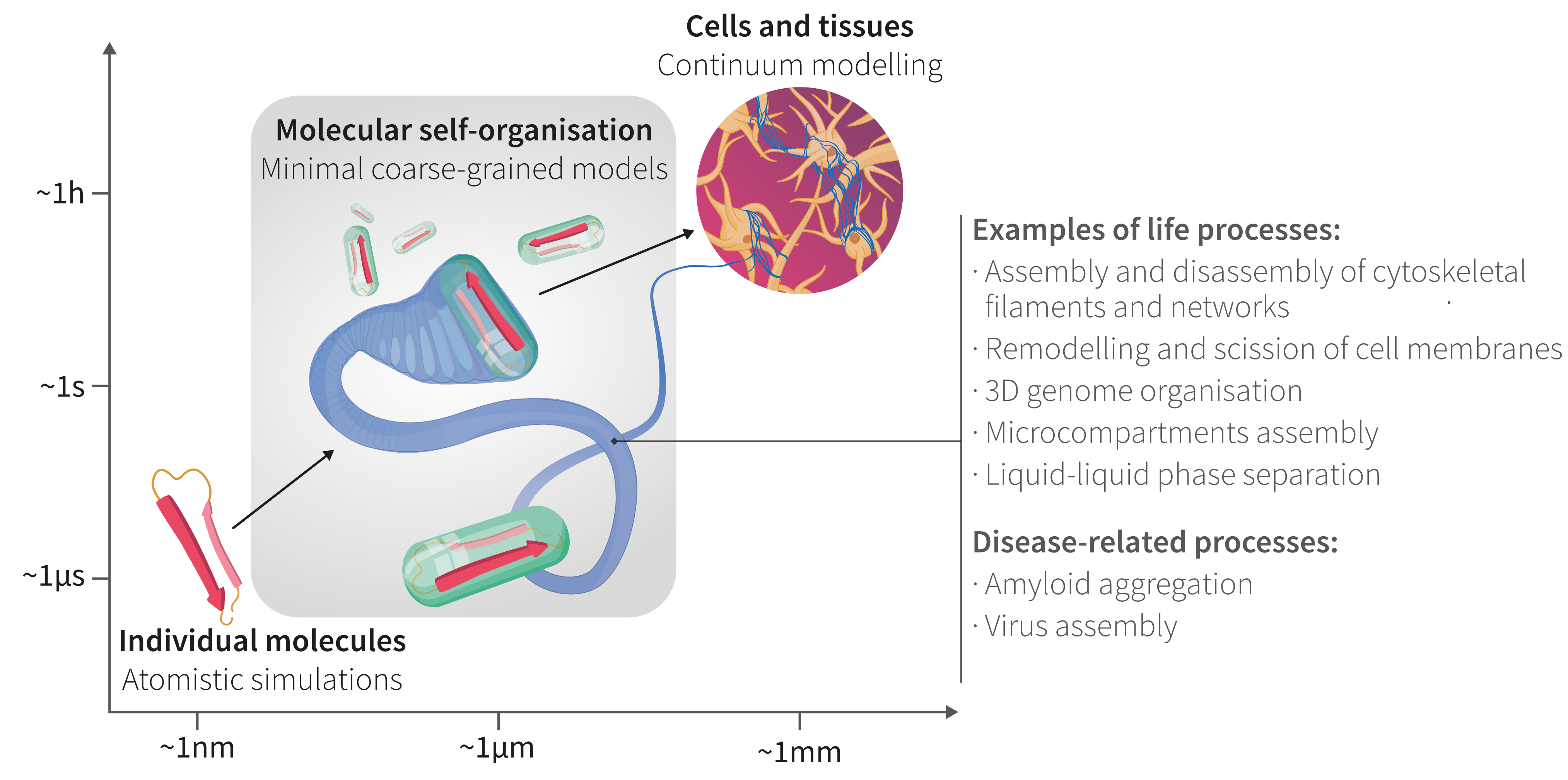}
	\caption{Minimal coarse-grained models, which account only for the crucial information on the shape and interactions of macromolecules, can be invaluable in bridging molecular and continuum scales, and identifying microscopic mechanisms behind life- and disease-related processes. The sketch represents the coarse-grained model developed for the assembly of  amyloid fibrils and extracellular plaques involved in neurodegenerative disorders~\cite{Saric2014, Saric2016}.}
\label{Figure0}
\end{figure*}

Due to their multiscale nature, macromolecular self-assembly phenomena are challenging to resolve in experiments, and are not amenable to atomistic molecular simulation. Luckily, to describe the assembly processes, not all molecular details are important, and the relevant physics can be efficiently captured by diffusive motion of macromolecules along the free energy landscape shaped by effective intermolecular interactions. The underlying physics can be represented via minimal coarse-grained models that retain only crucial information about the shape and interactions of the molecules, as sketched in Fig.~\ref{Figure0}. Because of their simplicity, such models can reach length and time scales comparable to those probed experimentally, and their results can  be often directly compared to data collected at macroscopic scales. Minimal coarse-grained models are hence invaluable in closing the gap between molecular and macroscopic scales --- the regime where the key physics  of the molecular machines of life occurs.  

In minimal coarse-grained models a macromolecule is typically described either by a single particle decorated with interaction patches, or as a collection of particles connected by springs. Figure 2 illustrates how such models are built. These models originate in statistical mechanics and soft matter physics where they have been traditionally used to describe collective behaviour of polymers, nanoparticles, and colloids~\cite{Schiller2018}. Unlike bottom-up coarse-grained models, which use a prescribed coarse-graining procedure to retain statistical accuracy or to preserve substantial information about sub-molecular details~\cite{Pak2018, Marrink2019}, the minimal coarse-grained models we describe here are at heart top-down. Instead of focusing on molecular details, they typically try to capture collective phenomena and the generality of the process under study.

These models contain a small number of parameters, hence it is easy to discern the influence of each parameter on the behaviour of the system. This makes it possible to identify the main driving forces in the system, analyse their roles in isolation from one another, and gain intuitive physical understanding of the biological phenomenon of interest. The key challenge lies in identifying the minimal set of parameters that will describe the experimentally observed behaviour, and make experimentally testable predictions. As there are no universal ``force-fields'' that can be used in this case, defining the appropriate (near) minimal set of parameters and exploring the parameter space can be a lengthy and non-trivial process.

However, these models can integrate a wealth of the available information from both the atomistic and macroscopic sides of the modelling spectrum. Examples include incorporating information from structural studies on the macromolecular shape or the position of interaction sites~\cite{ Perlmutter2013, Saric2014,Fusco2016,Paraschiv2019}, parametrising interactions to reproduce experimentally measured kinetic constants and binding constants or aggregate size distributions~\cite{Saric2014, Dear2018, Davis2019}, and parametrising the model to reproduce the shapes of assemblies observed in microscopy data~\cite{Pannuzzo2018,Harker-Kirschneck2019}. Although the absolute numerical values of the model parameters are not necessarily uniquely defined, the relative ratio of the important parameters needs to be preserved to give rise to the experimentally observed behaviour. 
Finally, the consumption of energy at  molecular scales, which typically occurs via binding or hydrolysing nucleotides that drive protein conformational changes, can be incorporated into the coarse-grained models by dynamically changing the molecular shapes/interactions or the relevant kinetic and binding constants.

In this review, we give a state of the art overview of the existing minimal coarse-grained models developed to describe molecular assembly in biology and its emergent functions. We focus on models that address biological phenomena, rather than self-assembly for materials, biomimetic, or therapeutic purposes. Since molecular self-assembly in living organisms is regularly coupled to an energy consumption, a more general term -- molecular self-organisation -- might be more appropriate~\cite{Wedlich-Soldner2018}. It is easy to notice that the coarse-grained models considered here cannot be easily sub-classified with respect to their ingredients, but rather with respect to the biological phenomenon that they focus on, leading to the "one system -- one model" pattern. This review is therefore organised according to the biological systems the models have been developed for, focusing mainly on the advancements achieved in the past 2--3 years.

\section*{Protein filament assembly}

\subsection*{Cytoskeleton}
The cytoskeleton provides a scaffold that shapes the mechanical characteristics of cells, drives membrane deformations and cell migration, and is involved in intracellular transport.
In order to form  complex networks that are adaptable to such specific cellular functions, the cytoskeletal filaments - actin and microtubules - undergo stochastic assembly and disassembly events. Although the assembly process of individual filaments has been intensively studied, especially for microtubules, open questions still remain.
In particular, the role of GTP hydrolysis, which leads to conformational changes in tubulin and is believed to trigger filament disassembly, and the impact of various microtubule associated proteins are still elusive. The microtubule instability has been  investigated with  dimer-scaled lattice models which share many common features~\cite{Hemmat2018,Vanburen2005,Margolin2012,Coombes2013,Zakharov2015,Piedra2016,Castle2017}. A microtubule is represented by a helical lattice composed of thirteen protofilaments, as shown in Fig.~\ref{Figure2}. Individual subunits can be added or removed with specific rates accounting for the growth and shrinkage of the filament. Lateral and longitudinal interactions between subunits may be incorporated, and the hydrolysis of GTP to GDP can be modelled by a change in the interaction strength or dissociation rate. While tubulin subunits are often assumed to arise from a population pool, free diffusion can also be explicitly considered using a Brownian dynamics approach~\cite{Castle2013}. Recently, such models were employed to further investigate the role of the GTP cap~\cite{Piedra2016}, the impact of microtubule-associated proteins on the dynamic instability of microtubules~\cite{Castle2017}, and the function of the prokaryotic cytoskeleton~\cite{Lan2009,Ghosh2008,Fischer-Friedrich2011}.

On larger length scales, cytoskeletal filaments self-organise into complex dynamic networks by the interplay between molecular motors, cross-linking proteins, and other binding proteins. 
How the interaction between these basic building blocks can assemble diverse dynamic network architectures is efficiently captured by coarse-grained computational models. These models often possess the same fundamental features, but vary when applied to specific biological systems. While many coarse-grained models for the cytoskeleton have been developed over the past decade~\cite{Nedelec2007,Vavylonis2008,Kim2009,Gordon2012,Popov2016,Freedman2017}, three prominent models for this class of simulations have been formulated as open-source computer simulation packages: Cytosim, developed by Ned\'{e}l\'{e}c et al.~\cite{Nedelec2007}, AFINES, introduced by Freedman et al.~\cite{Freedman2017}, and MEDYAN, developed by Popov et al.~\cite{Popov2016}. The first two are based on Langevin dynamics, whereas the third one uses a compartment-based reaction-diffusion scheme. These diverse approaches allow for various forms of mechano-chemical feedback between cytoskeletal filaments and their binding partners.

\begin{figure*}
	\centering
	\includegraphics[width = \linewidth]{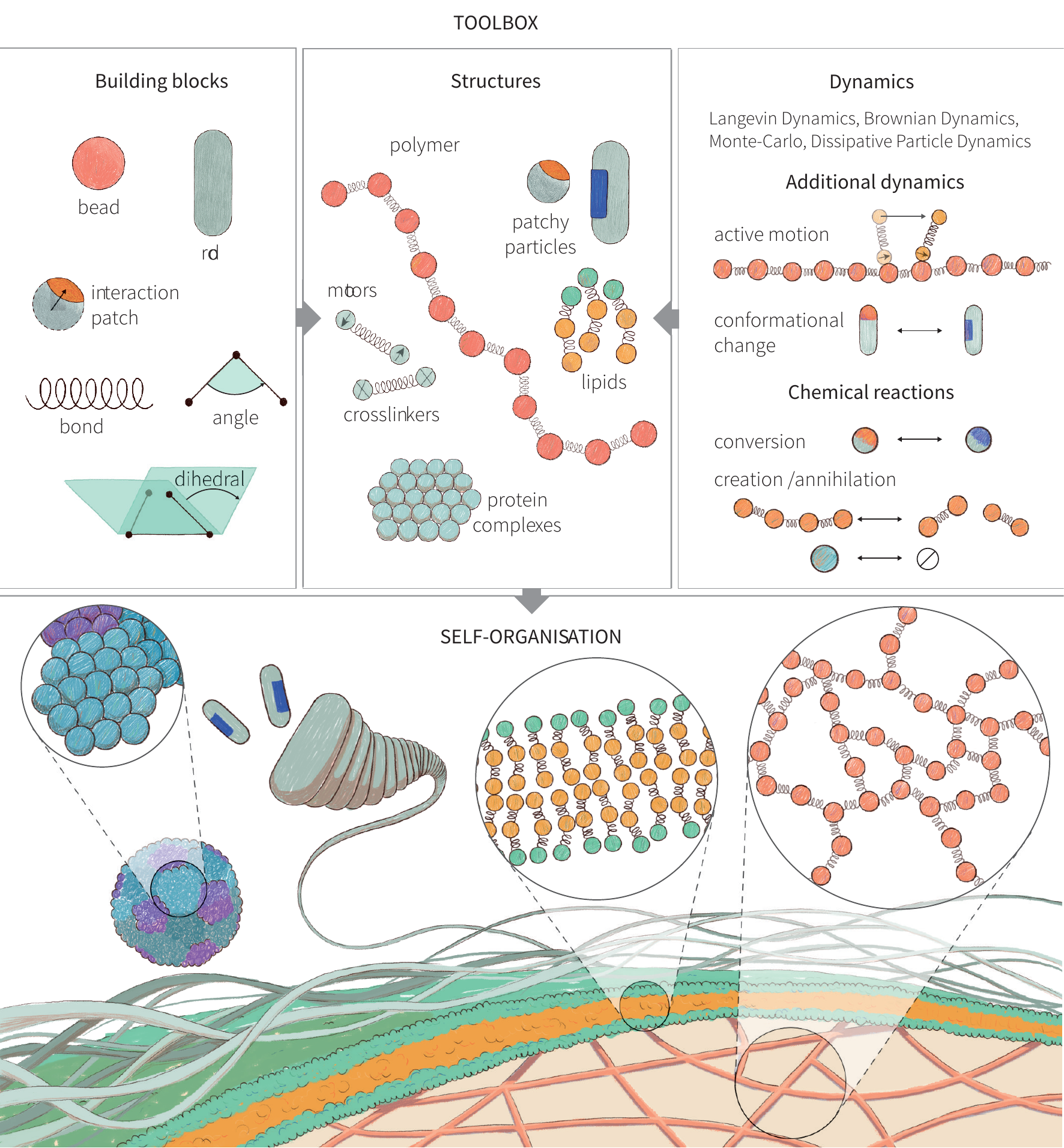}
	\caption{Building minimal coarse-grained models. The models are developed by connecting individual building blocks into structures using bonding and angular potentials (left panel). Specific interactions are often included by decorating the building blocks via patchy interaction sites. The structures (middle panel) are then simulated using the integration methods shown in the right panel, in which, depending on the system, additional dynamics or chemical reactions can be incorporated. This approach is then used to study the formation and function of cellular nanomachinery and related pathological assemblies as illustrated in the bottom panel. From left to right: virus capsid, amyloid fibrils, cellular membranes, and cytoskeleton.}
\label{Figure1}
\end{figure*}

In general, cytoskeletal filaments are modelled as dynamic worm-like chains. 
Both molecular motors and cross-linkers are taken into account by freely diffusing elastic springs which can bind to and unbind from filaments. 
Cross-linkers are immobile, whereas motors actively walk in either direction of the filament they attach to according to a linear force-speed relation, as sketched in Fig. \ref{Figure1}. 
Within such frameworks, Belmonte et al.~\cite{Belmonte2017} and Freedman et al.~\cite{Freedman2017} recently investigated the contractility of disordered cytoskeletal networks. Freedman et al.~\cite{Freedman2018b} showed that distinct actomyosin architectures arise in response to filament length and concentration, and the kinetics of motors and cross-linkers (Fig.~\ref{Figure2}). Stam et al. investigated~\cite{Stam2017} the impact of the filament rigidity and connectivity on the deformation modes of networks. While motor-driven semiflexible filaments mainly lead to contractile networks, networks of rigid filaments can be tuned from contractile to extensile networks through cross-linking. For cell division, Roostalu et al.~\cite{Roostalu2018} showed that both the ratio of motors to microtubules, and the ratio between microtubule growth speed and motor speed determine whether a microtubule architecture is polar or nematic. Furthermore, Blackwell et al. demonstrated that the formation and maintenance of antiparallel microtubule overlaps is essential for the assembly and stability of bipolar mitotic spindles in fission yeast~\cite{Blackwell2017}. Other recent applications range from the role of actin disassembly in chromosome transport in starfish oocytes~\cite{Bun2018}, and the role of cross-linkers in cytokinetic ring closure during cytokinesis of C. elegans~\cite{Descovich2018}, to the importance of actin nucleation for endocytotic processes~\cite{Mund2018}.


\subsection*{Formation of amyloid aggregates}
When proteins which are not meant to aggregate form assemblies, severe pathologies can develop. A pertinent example is the assembly of normally soluble proteins into amyloid fibrils, which is implicated in a range of human diseases, including Alzheimer's and Parkinson's disease~\cite{Dobson2019}. These processes typically occur at low protein concentrations, as low as nanomolar, and at long time-scales (hours \textit{in vitro} and years \textit{in vivo}). 
The conformational change from soluble to the fibril-forming structure is at the heart of these processes. Intermediate-resolution coarse-grained models previously developed by Shea, Caflisch, and Vendruscolo offered invaluable insights into the mechanisms of aggregation, as reviewed in~\cite{Wu2011,Ilie2019}, but were still computationally too expensive to reach physiological protein concentrations and experimentally-relevant time-scales.



In  recent years, \v{S}ari\'c et al.~\cite{Saric2014,Saric2016a}, building on work by Vacha et al.~\cite{Vacha2011}, developed a Monte-Carlo simulation model to capture the process of amyloid nucleation in solution. In this model the soluble protein conformation is represented by a hard spherocylinder with an attractive end patch that facilitates the formation of small non-fibrillar oligomers, similar to those observed in \textit{in vitro} experiments. The proteins can undergo a conformational change into their fibril-forming state, which is modelled by interchanging the attractive interaction tip with an attractive side patch, as sketched in Fig.~\ref{Figure1}. The model showed that the formation of amyloid fibrils at physiological concentrations proceeds via a two-step mechanism through transient oligomeric structures. The model was then extended to capture fibril self-replication~\cite{Saric2016}, showing that the adsorption of proteins onto the surface of pre-existing fibrils plays a crucial part in determining the rate of fibril self-replication. Importantly, the model enabled a  direct comparison with the kinetic measurements of  self-replication of A$\beta$ fibrils, and made mechanistic predictions that were subsequently confirmed in experiments~\cite{Saric2016}. The model has been used since to explain the experimentally-observed temperature-dependence of amyloid nucleation and self-replication~\cite{Cohen2018,Michaels2018a}, as well as reaction orders of amyloid aggregation reactions~\cite{Saric2016a,Meisl2017,Michaels2018}.

In a similar approach, Ilie et al. developed a Brownian dynamics model for amyloid aggregation of $\alpha$-synuclein~\cite{Ilie2016,Ilie2017}. Here, the soluble conformation of the protein is modelled as a soft spherical particle whereas the fibril-forming conformation is modelled as a spherocylinder with an attractive side patch. The authors highlighted the importance of the two-step nucleation mechanism and pointed out the possibility of a fibril self-replication cycle.  Finally, Lu et al. recently developed a three-bead  patchy particle model to investigate the structural and mechanical properties of amylin and A$\beta$ fibrils~\cite{Lu2018}. 


\begin{figure*}[t]
	\centering
	\includegraphics[width = \textwidth]{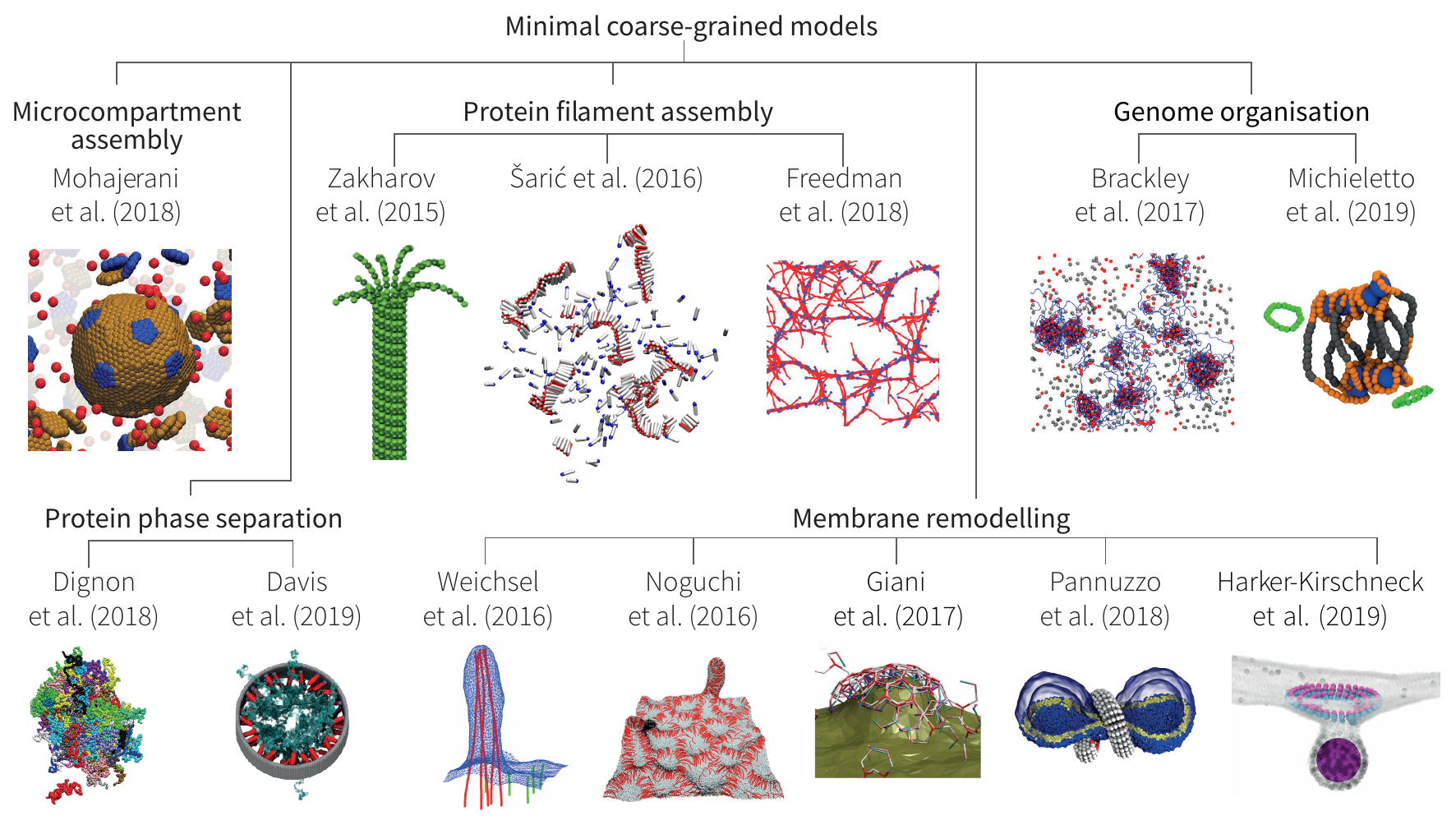}
	\caption{Overview of the systems and representative coarse-grained models discussed in this review. Individual snapshots reprinted with permission from~\cite{Mohajerani2018, Zakharov2015, Saric2016, Freedman2018b, Brackley2017b, Michieletto2019, Dignon2018, Davis2019, Weichsel2016, Noguchi2016, Giani2017,Pannuzzo2018, Harker-Kirschneck2019}.}
\label{Figure2}
\end{figure*}
\section*{Spatiotemporal genome organisation}
Investigating the biophysical mechanisms behind the spatial organisation of genetic material in the cell nucleus has recently emerged as a very active field of research. 
In the cellular environment a multitude of proteins interact with and bind to chromatin fibres to drive their spatial and structural organisation. In the context of minimal coarse-grained models, chromatin is prominently modelled as a flexible polymer chain, while proteins are spherical particles that bind to the chain and generate bridging-induced attractions~\cite{Brackley2013}. Post-translational modifications of DNA-binding proteins can also be included by modifying their affinity to DNA~\cite{Tootle2005, Brackley2017b}. Moreover, the bridging proteins can switch between active and inactive states at a pre-specified rate, which has been found to regulate the micro-phase separation of DNA-protein clusters~\cite{Brackley2017b}, see Fig.~\ref{Figure2}.

Buckle et al. further developed the polymer model for studying the three-dimensional chromatin organisation by parametrising it using widely available epigenetic and protein binding data~\cite{Buckle2018}. Consequently, the model returned better predictions of locus conformations for specific histone modifications. Michieletto et al. used a similar polymer model to investigate the emergence of epigenetic patterns~\cite{Michieletto2018}. 
In this model post-translation modifications and histone displacements are described by changing the properties of the polymer beads, which has been found to quantitatively reproduce the histone mark distribution appearing in Drosophila cells. A similar model was then used to understand the basic physical principles behind retroviral integration~\cite{Michieletto2019}, where integration events are accounted for by stochastic recombination to form bonds between the viral and host polymer chains, as shown in Fig.~\ref{Figure2}. 
Similarly, Fudenberg et al.~\cite{Fudenberg2016} studied chromatin organisation that occurs due to multiple cooperating loop extrusions, which explained diverse experimental data on the interaction patterns between different loci on the genome.

\section*{Assembly of protein microcompartments}
The genetic materials of viruses are encapsulated in a protein coat, called capsid.
Minimal coarse-grained molecular dynamics simulations have been widely used to investigate viral capsid formation as reviewed in Refs.~\cite{Hagan2014, Hagan2016}, and to investigate the role of nucleic acids in virus assembly~\cite{Perlmutter2013}. The capsid subunits, capsomeres, are typically described as a collection of beads that reproduces the overall shape of the subunit found in structural studies, as illustrated in Fig.~\ref{Figure1}. The charges and interaction patches on the capsomere were also chosen according to the structural data. The particles on different subunits interact through pairwise potentials which drive capsid assembly to a minimal energy state. 

Recently, modified models of virus assembly were also applied to the assembly of bacterial microcompartments~\cite{Mohajerani2018}, as shown in Fig.~\ref{Figure2}, and to the formation of capsids around differently-shaped nanoparticles~\cite{Lazaro2018,Zeng2018}. 
Coupling the models for capsid assembly with membranes, Shubertova et al. showed that virus entry can be prevented by multivalent inhibitors that block the receptor binding sites on the virus~\cite{Schubertova2017}, while  L\'{a}zaro et al. demonstrated that capsid cores guide the morphology of the enveloped virus~\cite{Lazaro2018a}. 


\section*{Remodelling of lipid membranes}
Cellular membranes are highly dynamic structures that are constantly subject to protein-induced shape changes. Although a wide variety of protein assemblies participate in sculpting and cutting cell membranes, the  majority of coarse-grained computer simulations focused on clathrin-mediated endocytosis~\cite{Daumke2014}. In the models discussed here the membrane is typically modelled using either a three-beads-per-lipid model~\cite{Cooke2005}, or highly coarse-grained options such as single-particle-layer membrane models~\cite{Yuan2010} and dynamically triangulated networks~\cite{Saric2013}. 

Based on the characteristic triskelion structure of the clathrin molecule, den Otter et al. developed a model in which the molecule is represented as a three-legged patchy particle interacting via a four-site anisotropic potential~\cite{denOtter2010}. The model is able to capture the assembly of polyhedral clathrin cages~\cite{denOtter2010} in solution, and when coupled to a triangulated membrane, it leads to the formation of a clathrin-coated pit~\cite{Giani2016, Giani2017}, as shown in Fig.~\ref{Figure2}. 


The formation of the clathrin coated pits can be assisted by curved proteins like BAR, which bind to the membrane surface inducing membrane curvature~\cite{Schoneberg2017a}. Simulations in which the BAR protein is described as a collection of beads arranged into a curved rigid body have demonstrated how membrane morphologies depend on the density and geometrical properties of the arc-shaped proteins~\cite{Bonazzi2019,Noguchi2016,Noguchi2017}, see Fig.~\ref{Figure2}. 

Finally, the scission of the membrane neck of the completed clathrin-coated pit is performed by the dynamin polymer. Recently, the interplay of constriction, elongation, and rotation in dynamin-facilitated membrane fission was elucidated with a coarse-grained simulation model~\cite{Pannuzzo2018}. Longitudinal rotation of the dynamin filament and the resulting torque on the lipid bilayer was found to play a pivotal role in mediating membrane fission at the tubular neck, as shown in Fig.~\ref{Figure2}. 


Endocytosis is an example of a process where the membrane is cut from the outer side of the membrane lumen. Recently, a  coarse-grained model of an ESCRT-III polymer, the only known machine that cuts the membrane from the inner side of the neck, has been developed by Harker-Kirschneck et al.~\cite{Harker-Kirschneck2019}. 
The main finding is that a change in the filament geometry from a flat spiral to a twisted helix guides the membrane deformation, while the return to the flat state drives the neck scission. The model is able to capture all the modes of ESCRT-III driven membrane deformations observed in a range of experiments, including upward and downward tubes and cones, as well as the formation of cargo-loaded vesicles. 

In one of the few examples of the coupling between cytoskeletal filaments and membrane deformations, Weichsel et al. studied the role of membrane properties in driving  cooperative bundling of actin filaments and the subsequent formation of actin-filled tubular protrusions~\cite{Weichsel2016}. In their model a triangulated membrane was coupled to a worm-like chain model of actin filaments, which were able to stochastically grow, shrink and protrude through the membrane. 

All of the above mentioned models consider the production of mechanical forces via protein assembly. Recently, a coarse-grained model for sensing of mechanical forces via mechanosensitive channels has been developed by Paraschiv et al.~\cite{Paraschiv2019}. The authors found that the assembly of channels into fluid-like clusters decreases the channel gating, exhibiting positive feedback between the protein assembly and their gating activity.
\section*{Functional phase separation}
Liquid-liquid phase separation is an important mechanism that facilitates spatiotemporal organisation of macromolecules in the cell. The field is rapidly evolving, and while most theoretical studies have used analytical tools, the number of simulations studies is on the rise too. Current research efforts focus on the link between the sequence-specific properties of proteins and their phase separation behaviour. Dignon et al. have developed a one-particle-per-residue polymer model taking into account the chemical properties of the individual residues, which enabled the authors to obtain phase diagrams of  proteins like Fused in Sarcoma, DEAD-box helicase protein LAF-1, and hnRNPA2~\cite{Dignon2018,Dignon2018a}, see Fig.~\ref{Figure2}.  Building  on this approach, Das et al. further studied the role of the sequence charge pattern in phase separation of intrinsically disordered proteins~\cite{Das2018}. Furthermore, Nguemaha et al.~\cite{Nguemaha2018} investigated the role of RNA and other regulators in the protein droplet formation, describing the protein-regulator system as a binary mixture of spherical particles with different interaction patch characteristics. 

Focusing on phase separation to facilitate chromatin organisation, Nuebler et al.\cite{Nuebler2018} determined that active loop extrusions and phase separation are two interdependent mechanisms that influence three-dimensional chromatin structures. In a recent work, the link between epigenetic patterns and genome organisation was also studied from the perspective of phase separation by Coli et al.~\cite{Coli2018}. 

Protein phase separation also appears to play a key role in the gating of the nuclear pore complex, a giant pore densely grafted by intrinsically disordered  proteins. Minimal coarse-grained models, which map one or two amino acids to a single bead, have been used to characterise the gel phase formation of the pore proteins~\cite{Ghavami2018}, as well as to show that the collective behaviour of nuclear pore proteins resides the crossover between the gel phase and an entropic brush phase~\cite{Davis2019}, see Fig.~\ref{Figure2}. 




\begin{tcolorbox}[enhanced,colback=orange!10!white,colbacktitle=orange!60!white,colframe=orange!80!white,coltitle=black,title={\textbf{Box: Chemical reactions} \label{Box}}]
Self-assembly processes in living organisms are often coupled to chemical reactions. A promising way to approach such complex processes is via combining particle-based reaction-diffusion frameworks with coarse-grained molecular dynamics simulations. This can be addressed by interacting-particle reaction dynamics, as introduced with the software package ReaDDy/ReaDDy2~\cite{Schoneberg2013,Biedermann2015,Hoffmann2018} in which chemical reactions are explicitly incorporated (Fig.~\ref{Figure1}). Event-driven techniques such as Green's function reaction dynamics~\cite{VanZon2005,VanZon2005b} and first-passage kinetic Monte-Carlo~\cite{Opplestrup2006,Donev2010}, which were originally developed for Brownian diffusion, are now also employed to accelerate complex interacting-particle reaction dynamics~\cite{Vijaykumar2015,Vijaykumar2017,Sbailo2017}. Even though such approaches are still at their infancy, their further development will be of crucial importance for the ability to capture the pathways by which molecular self-assembly occurs in living systems. 
\end{tcolorbox}

\section*{Conclusions}
The minimal coarse-grained models described in this review have enabled mechanistic understandings of the formation and biophysical function of a range of functional and pathological intra- and extracellular assemblies. Even though these models are often referred as "toy models", it is important to keep in mind that every model is an approximation suited for the scale and phenomena it addresses. This is true both for the models described here, as well as for continuum models, atomistic force fields, or sophisticated electronic structure methods. Despite their simplified appearance,  the models presented here are not cartoons: they preserve the correct physics, appropriate for the length- and time-scales they target.

To retain their minimal nature and be able to reach experimentally relevant time and length scales, such models are typically system specific. Interestingly, in the case where different groups independently developed models for the same system
the final models end up looking remarkably similarly, as evident from the models dealing with the assembly of cytoskeletal filaments and networks. This nicely demonstrates that some well-defined physical principles need to be preserved in each system, limiting the number of possible model designs, akin to convergent evolution in biology.

The models developed thus far typically focused on one system in isolation from the other structures in its environment. To approach biological complexity and function, future developments will necessarily need to start combining different structures, for instance dynamic cytoskeletal networks with cell membranes and membrane-bound macromolecules. Recent rapid developments in imaging techniques, including super-resolution microscopy, hold a great promise to drive parametrisation of coarse-grained models and validate their results~\cite{Schermelleh2019}.

Importantly, self-organisation in living organisms commonly occurs in non-homogeneous environments and involves chemical reactions of the assembling macromolecules. Incorporating chemical reactions within self-assembly processes will be crucial to correctly capture the function, turnover, or maturation of biological structures. Methods to account for this are currently under intense development (see Box 1 for an overview). In addition to chemical reactions, it will also  be of key importance to start incorporating other non-equilibrium driving forces that modulate the assembly processes in cells, such as the presence of mechanical forces and chemical gradients. We believe that these will be directions in which the development of future models is heading. 

\section*{Acknowledgements}
We acknowledge funding from EPSRC (A.E.H. and A.\v{S}.), the Academy of Medical Sciences (J.K. and A.\v{S}.), the Wellcome Trust (J.K. and A.\v{S}.), and the Royal Society (A.\v{S}.). We thank Nikola Ojkic and Shiladitya Banerjee for critically reading the manuscript, and Claudia Flandoli for helping us with figures and illustrations.


\section*{References and recommended reading}
\renewcommand\bibpreamble{Papers of particular interest, published within the period of review, have been highlighted as: 
\\

\textbullet \quad of special interest
\\
\textbullet\textbullet \; of outstanding interest

}


\end{document}